\documentclass{article}

\usepackage{amsmath, amsthm, amssymb}
\usepackage{algorithm}
\usepackage{algorithmic}
\usepackage{array}
\usepackage{mathrsfs}
\usepackage{amssymb}

\usepackage{graphicx}

\usepackage{bm}

\usepackage{tabularx,ragged2e,booktabs,caption} % fortable caption and title
\usepackage{epstopdf}% To incorporate .eps illustrations using PDFLaTeX, etc.
\usepackage{subfigure}% Support for small, `sub' figures and tables

\begin{document}

%\jvol{00} \jnum{00} \jyear{2015} \jmonth{January}

%\articletype{MANUSCRIPT}

\title{\textit{Hit and Run ARMS: Adaptive Rejection Metropolis Sampling with 
Hit and Run Random Direction} %\LaTeX\ guide for authors (Style 2 + NLM reference style)
}

\author
{Huaiye Zhang\\
Department of Statistics, Virginia Polytechnic\\
 Institute and State University, Blacksburg, VA 24061, USA\\
Yuefeng Wu$^*$\\
Department of Mathematics and Computer Science,\\ University
of Missouri Saint Louis, Saint Louis, MO 63121, USA\\
Lulu Cheng and Inyoung Kim\\
Department of Statistics, Virginia Polytechnic\\
 Institute and State University, Blacksburg, VA 24061, USA\\
% \received{February 2015}
}

\maketitle

\begin{abstract}
An algorithm for sampling from non-log-concave multivariate distributions is
proposed, which  improves the adaptive rejection Metropolis sampling (ARMS) 
algorithm by incorporating the hit and run sampling. It is not rare that the 
ARMS is trapped away from some subspace with significant probability in the 
support of the multivariate distribution.  While the ARMS updates samples only 
in the directions that are parallel to dimensions, our proposed method, the hit 
and run ARMS (HARARMS), updates samples in  arbitrary directions determined by 
the hit and run algorithm, which makes it almost not possible to be trapped in 
any isolated subspaces. The HARARMS performs the same as ARMS in a single
dimension  while more reliable in multidimensional spaces. Its performance  is
illustrated by a Bayesian free-knot spline regression example. We showed that 
it overcomes the well-known `lethargy' property and decisively find the global 
optimal number and locations of the knots of the spline function.

\end{abstract}

{\bf Keywords}: 
Adaptive Rejection Metropolis Sampling; Hit-and-Run Algorithm;  Regression
Splines;  Free-knot Splines, Empirical Bayesian Method.\\

{\bf Classcode}62G08, 62C12.\\

\section{Introduction}

Adaptive reject Metropolis sampling (ARMS) is a combination of adaptive rejection sampling (ARS) \cite{Gilks92} and a Metropolis-Hastings sampling \cite{Hastings70}. ARS was proposed for sampling from univariate
log-concave conditional distributions. As an extension of ARS, ARMS
\cite{Gilks95} removes the log-concavity restriction on the simulated
probability density functions. When the density to be sampled from is
multidimensional, a straight-forward approach is to embed ARMS within a Gibbs 
sampler, provided that all one-dimensional conditional densities can be
simulated by ARMS.  This type of approaches have been used widely in various 
areas since the first paper on ARMS \cite{Gilks95}. For some more examples,  
\cite{Meyer99}, \cite{Meyer00},\cite{Puter02}, \cite{Maranda13},   and 
\cite{Yelland14} are all using ARMS in Gibbs sampler for multivariate 
distributions. However, if the probability density function is multimodal  in a
multidimensional space, Gibbs sampler can be easily trapped around some of  the
modes indefinitely, resulting in inefficient and even unreliable samples.  In
section 2, We showed that applying Gibbs sampler with ARMS algorithm  to sample 
from a 2-dimensional distribution is trapped around 3 modes, while the target 
distribution has total 4 similar modes and is fairly smooth. Although, in
practice,  there are methods, such as using different proposal distributions, 
to prevent Gibbs or Metropolis-Hastings sampler being trapped in some subset of
the  support of  the target distribution,  whose probability is
distinctly  less than 1. These methods usually implemented by  trial and error.
They may still give impressions of convergence while  some important aspects of 
the target distributions are missing. Hence, there is no guarantee from them
that the  samplers have sampled from all the subspaces around (almost) every
local modes,  especially sampling from multidimensional spaces and the locations
of the modes are  unknown priori.

While Metropolis-Hastings algorithm may be stuck in only part of the support of
the distribution, applying ARMS to sampling from one-dimensional distribution
does  not generally have this problem, since it is an accept-reject method with 
an adaptive instrumental density function.
%?and?
In multidimensional situation, it is mainly due to the approach  through Gibbs
sampler that the ARMS can be trapped in some distinctly less than  1 probability
subset of the support of the target distribution. Particularly, we  believe that
such being trapped disadvantage is due to that the Gibbs sampler updates the
multidimensional samples in a fixed order and only in one dimension for  each
step. Note that Gibbs sampler does not necessarily sample in each dimension in
every step, it could sample some dimension fewer times than the others and
could sample in different order of the dimensions. Such alternative sampling
scheme may relieve the being trapping issue at some level.  However, such
alternative schemes are not standard and need to be done by trial and fail. 
Therefore, again, this is not a very reliable way to solve the being trapped 
issue of ARMS in multi-dimensional distribution sampling.

We propose an algorithm, which incorporate hit and run method and ARMS, instead 
of embedding ARMS in Gibbs sampler, to sample from general multivariate 
distributions. Hit and Run algorithm was introduced by \cite{Schmeiser80} and
\cite{Rubinstein82}
%See Robert L. Smith's 1996 paper
Its property has been studied by \cite{Andersen07} and \cite{Lovasz03}. Also,
hit and run and Gibbs samplers were compared by \cite{Chen93}.  Although there's
no theoretical result, the empirical study showed that the hit and run method 
estimators have less bias and standard errors than the corresponding Gibbs 
sampler estimators. The authors suggested that this is due to the less
autocorrelations  of the hit and run method than the Gibbs sampler. Plus,  since
hit and run method generates uniformly distributed directions , so it suffers 
the problem of being stuck in only part of the support of the distribution much 
less likely. Due to the difficulty of generating a signed distance for the hit
and  run method, griddy, acceptance/rejection, or Metropolis methods have been
used  and the resulting algorithms have been studied. These methods more or less
scarifice efficiency. In this paper, ARMS algorithm is used for generate the
signed distance. The resulting algorithm has not been proposed and studied 
explicitly in literature based on the authors' best knowledge. Combining the 
reliability of hit and run algorithm and the efficiency of ARMS, we can avoid 
the unreliable and inefficient issue due to the Gibbs sampler.

It is worth to mention that the hit and run is just one of the algorithms that 
deals with direction sampling. For example,  adaptive direction
sampling (ADS) was introduced in \cite{Roberts94}, where the authors  mentioned
the  Gibbs sampler has much slower convergence rate in high dimensional
situation  compared to the ADS. Incorporating ADS and ARMS looks promising and 
worth to be studied and compared with the hit and run method with ARMS, but is
out of  the scope of this paper.

As an example of the application of HARARMS, it was applied to solve a free
knots regression spline problem. Spline function is widely used in the analysis 
of two-dimensional data $(y_i, x_i)$. Early papers \cite{Rice69} \cite {Smith79}
have addressed the importance of the role of splines in smoothing and regression
analysis. Ruppert et al. A method of equally spaced knots was proposed in
\cite{Ruppert03}. This knot placement method is simple and straightforward to 
implement. However, the  nature of this method does not guarantee that knots are
placed at all critical locations, at which the underlying regression function
possesses  sharp changes. The other type of methods is called curve-fitting with
free-knot splines, where the number of knots and their locations are determined
from  the data. To discuss the full benefits of the spline approach, free-knot
methods  become an important and difficult problem.

Traditional methods \cite{Rice69} for the optimal knot selection is to add knots
in intervals where the residuals are inadmissible large. In stead of parameters,
the positions of the knots can be taken as the selection of functional type in
an  ordinary curve fitting problem \cite{Wold74}. The knots should be chosen as
to correspond to the overall performance of the data fitting. A new knot is
assigned  until the sum of squared residuals gets to a minimum. The recent 
literature propose several approaches to automatic knot selection that needs  to
search all possible models. Many of them are based on stepwise regression ideas.
For example, \cite{Smith82}, \cite{Friedman89}, \cite{Stone97}.
A stochastic search method was proposed in \cite{Spiriti13} for optimizing the
knot locations by using a continuous genetic algorithm. Bayesian methods for
fitting  free-knot splines have been considered in some literatures
 \cite{Smith96}, \cite{Denison97}, \cite{Dimatteo01},and
 \cite{Lindstrom02}.
These approaches employ continuous random search methodology through the use  of
Monte Carlo Markov chain algorithms although the objective is not minimization 
of Equation. A review and comparison of some of these approaches is given by
\cite{Wand00}.

Although most of the automatic knot selection procedures mentioned here have 
exhibited good performance, they all face a nonlinear problem with a lethargy 
property \cite{Jupp78}. The "lethargy" property is intrinsic to free-knot spline 
problems. In the other word, there may be local minima, saddle points, or even 
local maxima for the optimal knot placement. The "lethargy" typically cause many
problems for the derivative-based optimization methodology and could lead to a poor estimation.

We show that the proposed HARARM algorithm has good properties for solving
global  optimization problems, in particular, finding the the numbers and 
locations of the free knot regression model efficiently and reliably. The
overall  procedure of the free-knot idea is investigated by simulated data sets.

The proposed HARARMS algorithm is described in detail in Section 2,  where a
brief description, which has more details than above, to the related algorithms 
are also given. Then by comparing and discussing the performances of sampling in  
multidimensional space by ARMS with Gibbs sampler and Hit and Run algorithm in 
Section 3, we show that HAEARMS is significantly better in multidimensional
situation,  especially there are multi-modes, which is the case in almost all
real  world problems.  In the last section, the HARARMS is used to fit the data 
to the free knots regression spline, and the performance is assessed.

\section{Algorithms}

In the Bayesian context, the objectives of the modeling are usually posterior 
distributions of the unknown parameters. High dimensional distribution samples 
can be generated from the Gibbs sampling \cite{Gelfand90} 
straightforwardly for Bayesian inference. At each iteration of the Gibbs
sampling,  the parameter is updated by sampling a new value from its full 
conditional distribution. The full conditional distribution of a parameter  is
its distribution conditional on the data and on the current values of all the other parameters.
%For a high dimensional distribution, methods for constructing full conditional distributions must be very efficient.

For the completeness, we introduce the adaptive rejection sampling(ARS) and the
adaptive rejection Metropolis sampling(ARMS) first. We define $\rm{x}$ to be the random 
variable to be sampled.

The ARS extends the rejection sampling \cite{Ripley87} by adaptively adjusting
the sampling distribution. For a basic rejection sampling method, it requires
that a sampling  distribution $g({{\rm x}})$ for a random variable ${{\rm x}}$
can be  easily drawn and $M g({{\rm x}}) \geq f( {{\rm x}})$ given a finite
constant $M$.  The rejection sample method is very useful for sampling a full
conditional  distribution since the integration of $f({{\rm x}})$ is not needed.
In practice, this algorithm involves an evaluation and potential rejection step
after a  sample is drawn from $g({{\rm x}})$. For most of the time, it  is a
challenging task to pick up an appropriate sample distribution $g({{\rm x}})$
and  the constant $M$ in order to reduce the ratio of the rejection.

The ARS \cite{Gilks92} is proposed to use a sequence of sampling distributions
$g_m({{\rm x}})$ defined by piecewise linear functions $h_m ({{\rm x}})$:
\begin{equation}
h_m({{\rm x}}) =\min[L_{l-1,l}({{\rm x}};S_m),L_{l+1,l+2}({{\rm x}};S_m)], \: {{\rm x}}_l \leq {{\rm x}} < {{\rm x}}_{l+1},
\end{equation}
where  $S_m = \{ {{\rm x}}_l, l = 0,\cdots, m+1\}$ is a set of points in the
support of $f$. $L_{lj}({{\rm x}};S_m)$ denotes the straight line through 
points $[{{\rm x}}_l,{\rm ln}f({{\rm x}}_l)]$ and $[{{\rm x}}_j,{\rm ln}f({{\rm
x}}_j)]$.  Defining $M_m = \int {\rm exp} ( h_m({{\rm x}}) ) d{{\rm x}}$, The
sampling distribution  is given by:
\begin{equation}
g_m({{\rm x}}) = \frac{{\rm exp}(h_m({{\rm x}}))}{M_m}.
\end{equation}
Now the algorithm of ARS is derived as the following:

%\begin{algorithm}
%\caption{Adaptive rejection sampling(ARS)}
% \label{ARS}
% \begin{algorithmic}
% \STATE Step 0, Initialize $M$ and $S_m$;
%\FOR {$s$=0 to $S$}
% \STATE Step 1, Sample ${{\rm x}}$ from $g_m({{\rm x}})$; sample $U$ from ${\rm Uniform}(0,1)$;
% \STATE Step 2,
%\IF{$U>\frac{f({{\rm x}})}{\exp\{h_m({{\rm x}})\}}$ }
%       \STATE {\{- rejection step:}
%       \STATE {set $S_{m+1}=S_m\cup\{{{\rm x}}\}$;}
%       \STATE {relabel points in $S_{m+1}$ in ascending order;}
%       \STATE {increment $m$ and go back to Step 1; \}}
%\ELSE
%      \STATE  \{- acceptance step:
%      \STATE  set ${{\rm x}}_{s}={{\rm x}}$, increment s and return ${{\rm x}}_s$.\}
%\ENDIF
%\ENDFOR
%\end{algorithmic}
%\label{alg:algorithm1}
%\end{algorithm}

\begin{itemize}
\item [0.] Initialize $m$ and $S_m$.
\item [1.] Generate ${\rm x}\sim g_m({\rm x}),  U\sim  \mathbb U_{[0,1]}$
\item [2.] If $U>\frac{f({{\rm x}})}{\exp\{h_m({{\rm x}})\}}$ ,  update $S_m$ to
$S_{m+1}=S_m\cup\{{\rm x}\}$; otherwise, accept ${\rm x}$.
\end{itemize}

For  univariate cases, the average of iterations of the ARS to accept one point
depends on the  initial $S_m$ and the target distribution $f$. The biggest
limitation  of ARS is to require $f$ to be log-concave since $h_m({{\rm x}})$
needs  to be an envelope of ${\rm ln}f({{\rm x}})$.  The ARMS \cite{Gilks95} was
proposed to deal with non-log-concave densities. The ARMS incorporates a
Metropolis-Hastings  algorithm step to ARS. The Metropolis-The Hastings
algorithm is briefly covered as the following:
%\begin{algorithm}
%\caption{Metropolis-Hastings algorithm}
% \label{MH}
% \begin{algorithmic}
% \STATE Step 0, Initialize starting value ${{\rm x}}_s$, and $s=0$;
%\FOR {$s$=0 to $S$}

% \STATE Step 1, sample ${{\rm x}}$ from $q({{\rm x}}|{{\rm x}}_s)$; sample $U$ from ${\rm Uniform}(0,1)$;
% \STATE Step 2,
%\IF{$U>{\min\{1, \frac{f({{\rm x}})q({{\rm x}}_s|{{\rm x}})}{f({{\rm x}}_s) q({{\rm x}}|{{\rm x}}_s)} \}}$ }
%       \STATE  \{ - rejection step:
%       %\STATE
%       {  set $  {{\rm x}}_{s+1}={{\rm x}}_s  $; } \}
%\ELSE
%      \STATE  \{ - acceptance step:
 %     %\STATE
%      set ${{\rm x}}_{s+1}={{\rm x}}$, increment $s$ and go back to Step 1. \}
%\ENDIF
%\ENDFOR
%\end{algorithmic}
%\label{alg:algorithm2}
%\end{algorithm}

\begin{itemize}
\item [0.] Initialize ${\rm x}_s$
\item[1.] Generate ${\rm x}\sim q({\rm x}|{\rm x}_s), \sim \mathbb U_{[0,1]}$.
\item [2.] If $U>{\min\{1, \frac{f({{\rm x}})q({{\rm x}}_s|{{\rm x}})}{f({{\rm x}}_s) q({{\rm x}}|{{\rm x}}_s)} \}}$, ${\rm x}_{s+1}={\rm x}_s$; otherwise accept ${\rm x}$.
\end{itemize}
The Metropolis-Hastings algorithm(MH) \cite{Hastings70} improved the
Metropolis algorithm \cite{Metropolis53} by generating samples from a
proposal distribution $q({{\rm x}}|{{\rm x}'})$. The performance of the MH
algorithm depends on the quality of the proposal distribution. Similar to other
MCMC  algorithms, samples from the MH algorithm may stay in a local maximum of
$f$ due to an unsuitable proposal distribution.

To carry out the ARMS algorithm \cite{Gilks95}, let $S_m=\{{{\rm
x}}_l;\,l=0,\cdots,m+1 \}$ denote a current set of points in ascending order,
%where ${{\rm x}}_0$ and ${{\rm x}}_{m+1}$ are the lower and upper limits of
% domain $D$ of $f({{\rm x}})$. 
For $1 \le l \le  m$, let $L_{lj}({{\rm x}};\,S_m)$ denote the straight line through points $[{{\rm x}}_l,{\rm ln}f({{\rm x}}_l)]$ and $[{{\rm x}}_j,{\rm ln}f({{\rm x}}_j)]$. A piecewise linear function $h_{l,l+1}({{\rm x}})=\max[L_{l,l+1}({{\rm x}}),\min\{L_{l-1,l}({{\rm x}}),L_{l+1,l+2}({{\rm x}})\}]$, where ${{\rm x}}\in ({{\rm x}}_{l},{{\rm x}}_{l+1})$. $h_m({{\rm x}})=\{h_{l,l+1}({{\rm x}});\,l=1,\cdots,m-1\}$. Let $h_{0,1}({{\rm x}})=L_{0,1}({{\rm x}})$ and $h_{m,m+1}({{\rm x}})=L_{m,m+1}({{\rm x}})$ define the first and the last piecewise linear functions, respectively. The sampling distribution is then $g_{m}({{\rm x}})=\exp\{h_m({{\rm x}})\}/M_{m}$, where $M_{m}=\int \exp\{h_{m}({{\rm x}})\}d{{\rm x}}$. We further define ${{\rm x}}_{cur}$ and ${{\rm x}}_{new}$ as the current value and new sample from $f({{\rm x}})$. ARMS starts from $h_{0,1}({{\rm x}})=L_{0,1}({{\rm x}})$ and construct $h_{i,i+1}({{\rm x}})$; we include a new point and relabel points; we then reconstruct $h_{l,l+1}({{\rm x}})$.
%\subsection{Multiple dimensional ARMS}
%We reserve the bold font ${\textbf x}$ for the objective random variable vector.
For multivariate distribution, we reserve the bold font ${\textbf x}$ for the 
objective random variable vector. Let $k$ denote the dimension of the variable
and $k=1,\cdots,K$. Assuming $f({{\rm x}}_{(k)}|{\textbf x}_{(-k)})$ is a target
univariate distribution to be selected, where ${\textbf x}_{(-k)}$ is the
parameters except ${{\rm x}}_{(k)}$.  For ease of notation we write $f({{\rm
x}})$ below instead of $f({{\rm x}}_{(k)}|{\textbf x}_{(-k)})$. The algorithm of
ARMS embedded in Gibbs sampler for multivariate distribution is described as the following:

%\begin{algorithm}
%\caption{Multiple dimensional adaptive rejection Metropolis sampling}
% \label{ARMS}
% \begin{algorithmic}
% \STATE Step 00: Initializing ${\boldsymbol {{\rm x}}}_{(-1)}$;
%\FOR {$s$=0 to $S$}

% \FOR {$g$=1 to $G$}
% \STATE Step 0, change notation $f({{\rm x}})\leftarrow f({{\rm x}}_{(g)}|{\boldsymbol {{\rm x}}}_{(-g)})$; Initialize %$M$ and $S_m$;
% \STATE Step 1, sample ${{\rm x}}$ from $g_m({{\rm x}})$; sample $U$ from ${\rm Uniform}(0,1)$;
% \STATE Step 2,
% \IF{$U>\frac{f({{\rm x}})}{\exp\{h_m({{\rm x}})\}}$ }
%       \STATE {\{- rejection step:}
%       \STATE {set $S_{m+1}=S_m\cup\{{{\rm x}}\}$;}
%       \STATE {relabel points in $S_{m+1}$ in ascending order;}
%       \STATE {increment $m$ and go back to Step 1;}\}
% \ELSE
%      \STATE  \{- acceptance step:
%      \STATE  set ${{\rm x}}_{A}={{\rm x}}$;\}
% \ENDIF
% \STATE Sample $U$ from ${\rm Uniform}(0,1)$;

%  \IF{$U > \min \Biggl\{1, \frac {f({{\rm x}}_{A})\min[f({{\rm x}}_{cur}),\exp(h_m({{\rm x}}_{cur}))]}{f({\rm %x}_{cur}), \min[f({{\rm x}}_{A})\exp(h_m({{{\rm x}}}_{A}))]} \Biggr\}$}
%       \STATE \{-Metropolis-Hastings rejection step:
%       \STATE set ${{{\rm x}}}_{(g)}={{{\rm x}}}_{cur}$;\}
% \ELSE
%      \STATE \{-Metropolis-Hastings acceptance step:
%      \STATE set ${{{\rm x}}}_{(g)}={{{\rm x}}}_{A}$;\}
% \ENDIF
%			\STATE increment $g$ until $g=G$, then set $g$=1 and increment $s$.
% \ENDFOR
%\ENDFOR
%\end{algorithmic}
%\label{alg:algorithm3}
%\end{algorithm}

\begin{itemize}
\item [] For $k=1$ to $K$, let $f({\rm x})$ denote $f({\rm x}_{(k)}|{\rm
x}_{(-k)})$ in the following steps. Do
\item [1.] Simulate ${\rm x}_{A}$ from $g_m({\rm x})$ and $U\sim \mathbb U_{[0,1]}$ until
$$
U\leq \frac{f({\rm x}_{A})}{\exp\{h_m({\rm x}_{A})\}}.
$$
\item [2.] Generate $U\sim \mathbb U_{[0,1]}$ and take\\
\begin{equation*}
{\rm x} _{(g)}=\left\{
\begin{array}{ll}
{\rm x}_{A} & \qquad U < \min \Biggl\{1, \frac {f({{\rm x}}_{A})\min[f({{\rm x}}_{cur}),\exp(h_m({{\rm x}}_{cur}))]}{f({{\rm x}}_{cur}), \min[f({{\rm x}}_{A})\exp(h_m({{\rm x}}_{A}))]} \Biggr\} ,\\
{\rm x}_{cur} &\qquad otherwise.
\end{array}
\right .
\end{equation*}
\end{itemize}

The Hit-and-Run algorithm can be thought of as random-direction Gibbs: in each
step of the hit-and-run algorithm, instead of updating ${{\rm x}}$ along one of
the dimensions, we update it along a randomly generated direction that is not
necessarily parallel to any dimension. More precisely, the sampler is defined in
two steps: first, choose a direction ${\rm d}$ from some positive density on the
unit sphere $d'd = 1$. Then, similar to Gibbs, sample the new  point ${{\rm
x}}_{new}$ along the line specified by $d$ and the distance by  ${\rm z}$, where
${\rm z}$ is from the marginal one-dimensional density on the line that
specified  by $d$. The ARMS with Hit-and-Run (HARARMS) algorithm is as the
following:

\begin{itemize}
\item [0.] Initiate ${\rm x}_s$, generate $d_g=u_g/\sqrt{u_1^2+\cdots + u_G^2}$, $u_g\sim \mathbb U_{[0,1]}$,and set $f^*(z)=f({\rm x}_s+d^Tz)$, initiate $M$ and $S_m$ for $f^*(z)$.
\item [1.] Simulate $z_{A}$ from $g_m(z)$ and $U\sim \mathbb U_{[0,1]}$ until
$$
U\leq \frac{f(z_{A})}{\exp\{h_m(z_{A})\}}.
$$
\item [2.] Generate $U\sim \mathbb U_{[0,1]}$ and take\\
\begin{equation*}
{{\rm x}} _{s}=\left\{
\begin{array}{ll}
 {{\rm x}}_{s-1}+d^T z_{A} & \qquad U < \min \Biggl\{1, \frac {f({z}_{A})\min[f({z}_{cur}),\exp(h_m({z}_{cur}))]}{f({z}_{cur}), \min[f({z}_{A})\exp(h_m({z}_{A}))]} \Biggr\} ,\\
{{\rm x}}_{s-1}+ d^T z_{cur} &\qquad otherwise.
\end{array}
\right .
\end{equation*}
\end{itemize}

%\begin{algorithm}
%\caption{Adaptive rejection Metropolis sampling with Hit-and-Run}
% \label{ARMSHR}
% \begin{algorithmic}
% \STATE Step 00, Initializing ${\textbf x}_s$, where $s=0$; Set $f^*({z})=f({\textbf x}_{s}+{\rm d}'{z})$;

%\FOR {$s$=1 to $S$}
% \STATE Step 0, generate $d_g=u_g/\sqrt{u_1^2,\cdots,u_G^2}$, $u_g \sim {\rm Unif}(0,1)$; Initialize $M$ and %$S_m$ for $f^*({z})$;
% \STATE Step 1, sample ${z}$ from $g_m({z})$; sample $U$ from ${\rm Uniform}(0,1)$;
% \STATE Step 2,
% \IF{$U>\frac{f^*({z})}{\exp\{h_m({z})\}}$\{}
%       \STATE {- rejection step:}
%       \STATE {set $S_{m+1}=S_m\cup\{{z}\}$;}
%       \STATE {relabel points in $S_{m+1}$ in ascending order;}
%       \STATE {increment $m$ and go back to Step 1;}\}
% \ELSE
%      \STATE  - acceptance step:
%      \STATE  set ${z}_{A}={z}$;
% \ENDIF
% \STATE Sample $U$ from ${\rm Uniform}(0,1)$, $z_{cur}=0$;

%  \IF{$U > \min \Biggl\{1, \frac {f^*({z}_{A})\min[f^*({z}_{cur}),\exp(h_m({z}_{cur}))]}{f^*({z}_{cur}), %\min[f^*({z}_{A})\exp(h_m({z}_{A}))]} \Biggr\}$}
%       \STATE \{-Metropolis-Hastings rejection step:
%       \STATE set ${\textbf x}_{s}={\textbf x}_{s-1}+{\rm d}'{z}_{cur}$ \};
% \ELSE
%      \STATE \{-Metropolis-Hastings acceptance step:
%      \STATE set ${\textbf x}_{s}={\textbf x}_{s-1}+{\rm d}'{z}_{A}$;\}
% \ENDIF
% \STATE increment $s$.
%\ENDFOR
%\end{algorithmic}
%\end{algorithm}

\section{Comparing sampling performances}
The ARMS with Hit-and-Run reduces a multiple dimensional question into one
dimensional ARMS sampling. It breaks the procedure into two steps: (1). pick up 
a random direction {\rm d}; (2). implement a one dimensional ARMS for the random
variable z. An important advantage of ARMS with Hit-and-Run over regular
multiple  dimensional ARMS is that it is much more likely to reach the  isolated
local areas by evaluating one dimensional ARMS in a random direction  searching,
while regular multiple dimensional ARMS only searches the space in the 
direction that is parallel to  one of dimensions  in each updating step. For
example, in a two dimensional case, the Gibbs sampler  sampling the new points 
either in the vertical or the horizontal direction from the current point. The 
following example shows that even for sampling from a mixture of 4 bivariate
normal  distribution with similar mixing probability mass and variances, the
Gibbs  sampler with ARMS can be trapped around only 3 of the 4 modes and totally
missing the  forth one. This is mainly due to that the forth mode is hardly
reached by searching in the 2 dimensional space along the directions only 
parallel to the axes in each step.

Assuming that we have a two dimensional random variable, ${\textbf x}=\{x_{(1)},x_{(2)}\}^{\rm T}$, and the the objective sampling distribution is $f({\textbf x})$ as below:
\begin{equation} \label{eqn:3.3}
f({\textbf x})=\sum_{i=1}^4{p_i{\rm N}({\textbf x}|\mu_i,\Sigma)}.
\end{equation}
In this example, we set $\mu_1=(5,-5)^{\rm T}$, $\mu_2=(5,5)^{\rm T}$, $\mu_3=(-5,5)^{\rm T}$, and $\mu_4=(13,13)^{\rm T}$. $\Sigma={\textbf Diag}(0.5,0;0,0.5)$, $p_1=0.2$, $p_2=0.3$, $p_3=0.2$ and $p_4=1-p_1-p_2-p_3$.

Sampling ${\textbf x}$ from $f({\textbf x})$ is an easy task if we know  that
$f({\textbf x})$ is a mixture of normal distributions. We can pick one sampling 
distribution among the four in terms of their probabilities $p_i$, and then
sample  ${\textbf x}$ from the selected normal distribution. Consider
$f({\textbf x})$  as if it did not have the explicit and easy form of mixture 
to sampling form, and treat it  as a general distribution, which is the case as 
sampling from general complex functions. Then we use both the Gibbs sampler with
ARMS embedded and the HARARMS to sample ${\textbf x}$ from the $f({\textbf x})$,
and compare their performances. After $s$=10000 iteration, we generate the
sampling  of ${\textbf x}$ from both  Gibbs sampler with ARMS embedded and
HARARMS.  To make a reference, contour plots are made by the grid search on 
$x_{(1)}$ and $x_{(2)}$, which refer as x1 and x2 in the Figure $\ref{fig:ARMS_gaussian}$.

{
\begin{figure}[!]
\includegraphics[width=\linewidth]{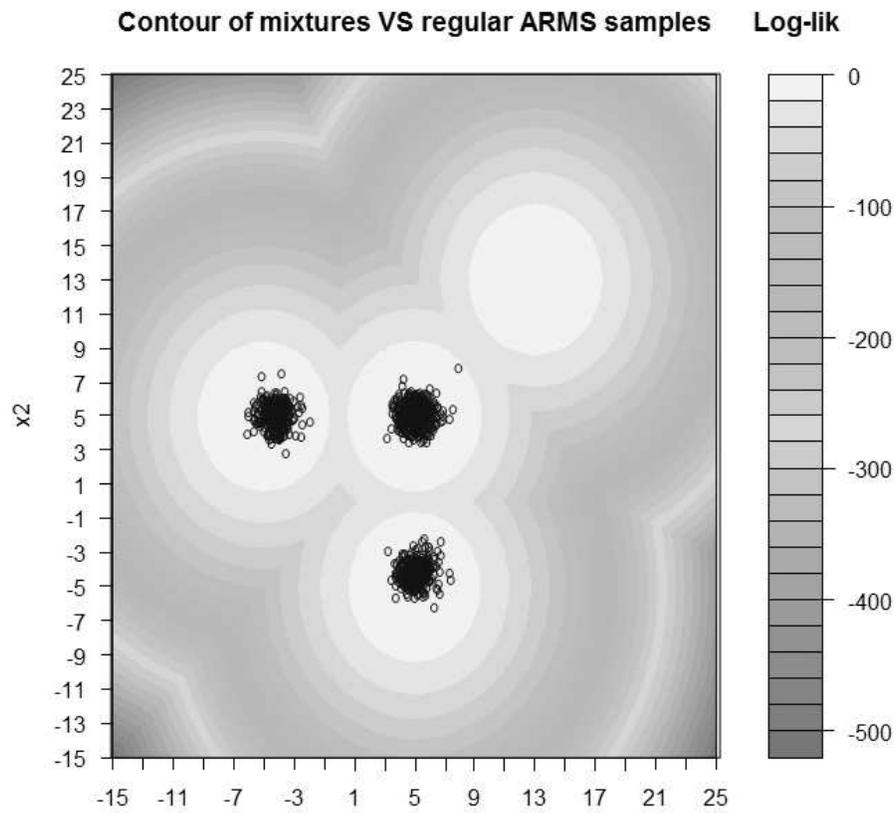}
\includegraphics[width=\linewidth]{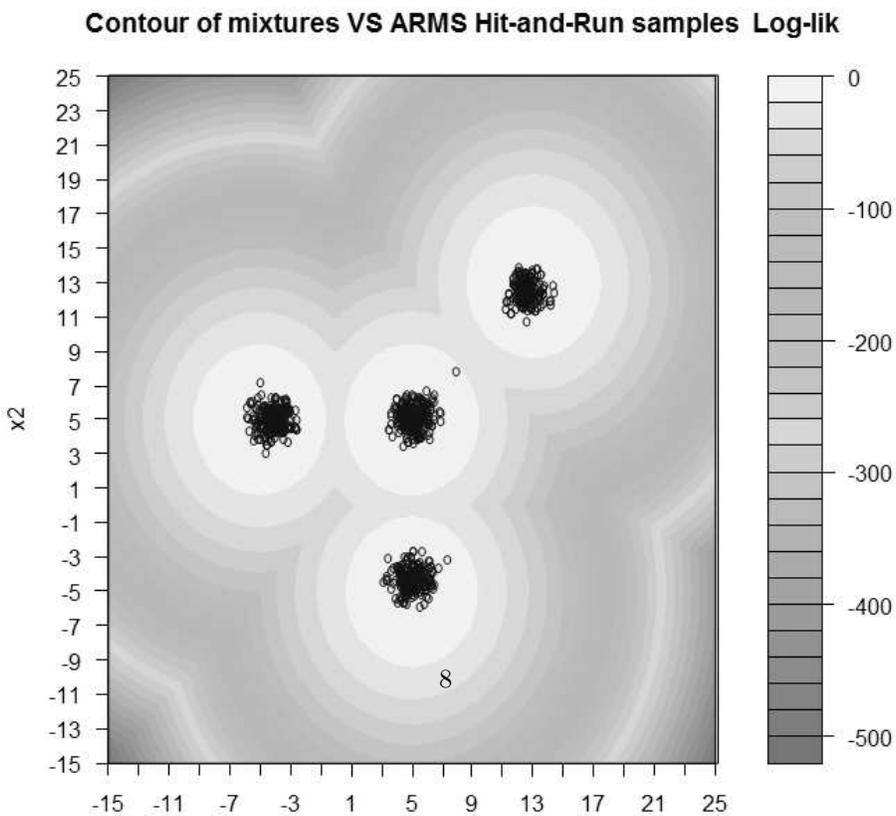}
\caption{   Top: shows that Gibbs-ARMS is trapped around only 3 of the 4 modes.
Bottom: shows that HARARMS samples from around every modes. }
\label{fig:ARMS_gaussian}
\end{figure}
}

The comparison of the performances of the two algorithms is also summarized  by the following table:\\

\begin{minipage}{\linewidth}
\captionof{table}{Comparing ARMS performances } % title of Table
%\centering  % used for centering table
{\begin{tabular}{ c|c|c|c|c|c|c|c|c|c} % centered columns (4 columns)
\hline\hline                      %inserts double horizontal lines
Parameter   & \multicolumn{3} {c|} {Actual} & \multicolumn{3} {c|} {HARARMS} & \multicolumn{3} {c} {ARMS with Gibbs} \\  % inserts table
\hline
      &   {$\mu_1$}     &  {$\mu_2 $} &{$p$} &    {$\mu_1$}   &      {$\mu_2 $}&{$p$}&   {$\mu_1  $}&{$\mu_2$ } &{$p$}\\  % inserts table
\hline
  (Comp 1)    &     {5 } &     {-5 }  & {0.20}   &    {5.00}      &    {-4.98} & {.20}  &  {4.99}      &   {-5.03} & {.29}    \\  % inserts table
  (Comp 2)    &    {5}   &    { 5 }   & {0.30}   &     { 4.99 }   &  {5.02}    & {.30}  & {4.99}       &   5.02    & {.42}      \\  % inserts table
  (Comp 3)    &     {-5} &      {5}   & {0.30}   & { -4.97 }      &    {4.99}  & {.22}  &  {-4.98}     &  4.99     & {.29}        \\  % inserts table
  (Comp 4)    &     {13} &        {13}&  {0.20}  &    {12.98}     &   {13.02}  & {.28}  &              &   &           \\  % inserts table
\hline
\hline
\end{tabular}}
{\small The HARARMS estimates all the parameters well, while the Gibbs sampler
misses one of the modes and gives severely biased estimation on the mixing probability.} \label{table:ARMS comparing1} % is used to refer this table in the text
\end{minipage}

\section{ Free Knots Regression splines}\label{sec:sec2}
Now we apply the HARARMS to a free knot spline model. Assuming that we would like to model a dependent variable, $y_i$, using one independent variable, $x_i$, where $i =1,\cdots,n$. Starting with the straight line regression model:
\[
y_i = \beta_0 +\beta_1 x_i + \epsilon_i,
\]
where $\epsilon_i \sim {\rm N}(0,\sigma^2)$. Denote $x=(x_1,\cdots,x_n)^{\rm T}$ and $y=(y_1,\cdots,y_n)^{\rm T}$. In the framework of spline regression, the basis functions for fitting such simple linear regression are defined as:
\[ \rm{X} = \left[ \begin{array}{cc}
1      & x_1 \\
\vdots  & \vdots \\
1      & x_n \end{array} \right].\]
To handle nonlinear structure, regression models consisting of several differently sloped lines are developed. One method is to introduce a basis function that is zero to the left of $\kappa$ and then is a positive value from $\kappa$ onward. A mathematical way of expressing this truncated basis function is
\[
(x_i - \kappa)_+,
\]
where, for any number $u$, $u_+$ is equal to $u$ if $u$ is positive and is equal to 0 otherwise. Therefore, the regression model with multiple truncated basis functions is
\begin{equation} \label{eqn:1.1}
y_i = \beta_0 +\beta_1 x_i + \sum_{k=1}^K b_k (x_i - \kappa_k)_+ +\epsilon_i, \hspace{0.3 in} i=1,\cdots,n.
\end{equation}
Now the corresponding basis function matrix is:\\
\begin{equation*}
\rm{X} = \left[ \begin{array}{cccccc}
1      & x_1    &  (x_1 - \kappa_1)_+ & \cdots &  (x_1 - \kappa_K)_+           \\
\vdots  & \vdots  &  \vdots              & \vdots &  \vdots                        \\
1      & x_n    &  (x_n - \kappa_1)_+ & \cdots &  (x_n - \kappa_K)_+
\end{array} \right].
\end{equation*}\\
In (\ref{eqn:1.1}) the value of $\kappa$ corresponding to the function
$(x_i-\kappa)_+$ usually refers to a knot. A set of such functions is called a
linear spline basis. Equation (\ref{eqn:1.1}) is called a spline model with a
linear spline basis. Assume that the parameters $\beta_0$, $\beta_1$, as well as $\textbf{b}=(b_1,\cdots,b_k,\cdots,b_K)$ are unknown while  $\kappa$'s in matrix ${{{\rm X}}}$ is known, the model can be easily solved by the multiple linear regression.

However, if the number and locations of the knots, the $\kappa$'s, are unknown, 
the model will be not so easy to handle, and the choice of them is a critical
problems.  Wold \cite{Wold74} describes the choice of knot positions as the
selection of functional type. The knots should be chosen in terms of the overall
performance of the data fitting. Simplified methods (\cite{Rice69},
\cite{Ruppert03}) works well when the number of knots is large (about 30-40).
The extra caution is necessary in using the large number of knots, since the 
great flexibility of splines can make overfitting the data. Ideally we should
put  as few knots as possible.  The knots should be placed at all critical
locations,  at which the underlying regression function possesses sharp changes.

The difficulty in detecting the optimal number and location of knots comes from
the  "`lethargy"' property (\cite{Jupp78}) intrinsic to free-knot splines as
discussed in Section 1. Such local maxima problem for free-knot splines method 
is rarely discussed, and it actually is the big barrier for designing an
attractive  free-knot algorithm. We show the local maxima problem for model 
(\ref{eqn:1.1}) in detail as the following:

To find the optimal number and locations of the knots, we define the loss 
function by the sum of squared errors, just the same as the classical method, 
which minimizes the loss function and finds the optimal. Let 
$\textbf{B}=(\beta_0,\beta_1,\textbf{b})^{\rm T}$. Based on (\ref{eqn:1.1}) and
assumption that the error is normally distributed, the full likelihood is:
\begin{equation} \label{eqn:1.3}
p(y_i|\kappa,\textbf{B},\sigma^2) = \frac{1}{\sqrt{2 \pi}\sigma}{\rm exp}[-\frac{1}{2}\sigma^2 (y_i-{{{\rm x}}_i(\kappa)}{\textbf B})^2]
\end{equation}
In (\ref{eqn:1.3}), both of knot locations $\kappa$ and the coefficients
$\textbf{B}$ and $\sigma^2$ need to be optimized. Given the number and locations of knots, the optimization in terms of the coefficients, \textbf{B} and $\sigma^2$, can be obtained by standard linear least-squares solution. In the other word, we have the  residual sum of squares for a given knots' locations by substituting $\textbf{B}$ and $\sigma^2$ with the ordinary least squares solution $\hat{\textbf{B}}$ and an unbiased estimation $\hat{\sigma^2}$. Consider an empirical Bayesian method:
\begin{equation} \label{eqn:1.4}
p(y_i|\kappa) \propto \frac{1}{\sqrt{2 \pi}\hat{\sigma^2}}{\rm exp}[-\frac{1}{2}\hat{\sigma^2} (y_i-{\rm X_i(\kappa)}{\hat{\textbf B}})^2]
\end{equation}
where $\hat{\textbf B}$ is an empirical solution, i.e.  $\hat{\textbf{B}}=({\rm
X}'{\rm X})^{-1}{\rm X}'y$. $\hat{\sigma^2} = \sum_i (y_i-{\rm
X_i}{\hat{\textbf{B}}})/(n-q-1)$,  where $q$ is the number of unknown
coefficients of (\ref{eqn:1.1}). It is now well known that the likelihood
function corresponding to (\ref{eqn:1.4}) tends to have multiple maxima and
exhibits a lethargy property near extremes that can lead to premature
convergence at  nonoptimal knot locations.

To show the multiple maxima of the likelihood function, we give a noninformative prior  to $\kappa$. Then the posterior distribution of $\kappa$ is:
\begin{equation} \label{eqn:1.7}
p(\kappa|y) \propto \prod_{i=1}^n p(y_i|\kappa).
\end{equation}
Therefore, the Log-likelihood function for $\kappa$ is
\begin{equation} \label{eqn:1.8}
l(\kappa,y) = \sum_{i=1}^n \log(p(y_i|\kappa)).
\end{equation}

Now we generate example data sets, and show the existence of the local  maxima
in terms of the location of the knots. Generate a simulated  Dataset 1, $(x_i,
y_i)$, from  (\ref{eqn:1.1}) with the parameters values as summarized in the
following Table \ref{table:nonlin1}, and the generated data is plotted in
Figure\ref{fig:data1}.

\begin{table}[!h] 
\caption{Simulated Dataset 1} % title of Table

\centering  % used for centering table
\begin{tabular}{c | c} % centered columns (4 columns)
\hline\hline                        %inserts double horizontal lines
Type & Value \\ [0.5ex] % inserts table
%heading
\hline
Knot location, $\kappa$ & $\kappa$=(200, 300, 400, 500, 700, 900) \\ % inserting body of the table
Coefficients, ${\textbf B}$ & $\beta_0 = -0.5$, $\beta_1 = -0.5$, and $\textbf{b} = (0.5, 1.0, -2.0, 2.5, -3.0, 3.5)$ \\
Error, $\epsilon_i$ & $\epsilon_i = N(0,30^2)$ \\
Independent variable, $x$ & $x=(1,2,\cdots,1000)^{\rm{T}}$ \\
Dependent variable, $y$&  Sampled based on (\ref{eqn:1.1})\\[1ex]      % [1ex]
% adds vertical space
\hline %inserts single line
\end{tabular}
\label{table:nonlin1} % is used to refer this table in the text
\end{table}

{
\begin{figure}[!h]
\includegraphics[width=\linewidth]{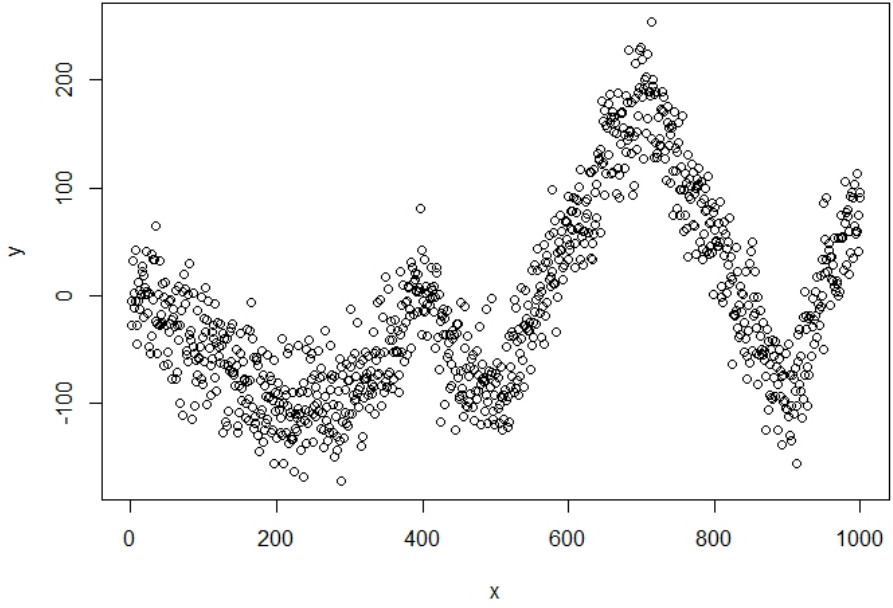}

\caption{ \baselineskip=12pt  $y$ is generated from (\ref{eqn:1.1}) with defined
knot locations and parameters values. $x=(1,2,\cdots,1000)^{\rm{T}}$
% Left: \baselineskip=12pt  Gridding the knot location with one knot splines, plot the log likelihood of \eqref{eqn:1.4} to demonstrate the multiple local maximum problem. Right: \baselineskip=12pt  %Gridding
%the knot location with two knot splines, plot the log likelihood of \eqref{eqn:1.4} to demonstrate the multiple local maximum problem.
} %\label{fig:figarms1}
\label{fig:data1}
\end{figure}
}

 Similar to the setting of dataset 1, let  $\kappa$ equal to $(700)$ or $(700,
 500)$, and then  plot the curve of the likelihood functions (\ref{eqn:1.8}) 
 versus the location of $\kappa$  in Figure \ref{fig:lik} correspondingly. Using
 one and two dimensional $\kappa$ is to make the graph easy to read, without
 loss  of generality. The multiple maxima are clear in the graph. Specifically, 
 for $K=1$, the largest value of the Log-likelihood (-5630.89) occurs with knot 
 location at 730; a local maxima occurs at 160, and the Log-likelihood is
 -5698.99.   For $K=2$,  Let both of $\kappa_1$ and $\kappa_2$ have values  from
 100 to 990 with increments of 10. The value of the Log-likelihood is then 
 plotted for each combination of the two knot locations on the right side of 
 Figure \ref{fig:lik}. You can see that the largest value of the Log-likelihood 
 (-5339.10) occurs with knot location at a splines function with $\kappa_1=670$ 
 and $\kappa_2=520$; A local maxima Log-likelihood (-5406.38) appears at a
 splines  function with $\kappa_1=700$ and $\kappa_2=270$; The other local 
 maxima Log-likelihood (-5535.50) appears at the combination of $\kappa_1=90$ 
 and $\kappa_2=770$. Through this visualization, It is clear that three local 
 Log-likelihood maximums exist within the splines function with two knots.   
 This indicates that the log-likelihood function for the general free knot 
 spline with unknown number  of knots  should be multimodal in the high 
 dimensional space.   we can see that the location of a global maximum can 
 therefore be assured only through a global grid search, which unfortunately 
 becomes computationally infeasible quickly as the dimension increases. As we 
 discussed above, for sampling from such target functions, the HARARMS will be 
 better than Gibbs sampler.

{\small
\begin{figure}[!h]
\includegraphics[width=.7\linewidth]{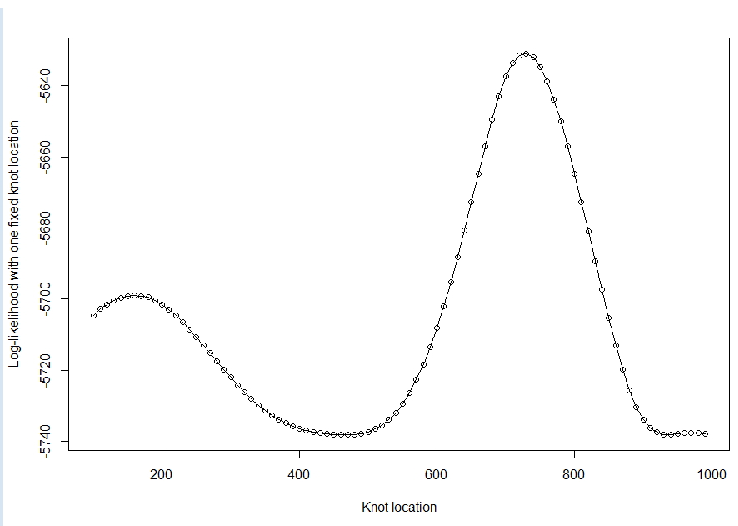}
\includegraphics[width=.7\linewidth]{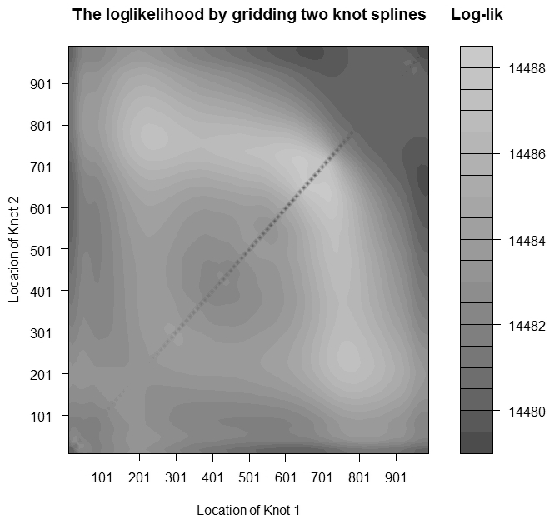}

\caption{% Left:  \baselineskip=12pt  $y$ is generated from \eqref{eqn:1.1} with defined knot locations and parameters values. $x=(1,2,\cdots,1000)^{\rm{T}}$
 Top: \baselineskip=12pt  Gridding the knot location with one knot splines, plot
 the log likelihood of (\ref{eqn:1.4}) to demonstrate the multiple local maximum
 problem.  
 Bottom: \baselineskip=12pt  Gridding the knot location with two knot splines,
 plot the log likelihood of (\ref{eqn:1.4}) to demonstrate the multiple local
 maximum problem. } \label{fig:figarms1}
\label{fig:lik}
\end{figure}
}

Table \ref{table:nonlin2} shows another setting for generating a quadratic 
spline function with 5 knots. The data generated from the this setting is named
as dataset 2.\\

\begin{table}[h]
\caption{Simulated Dataset 2 - for Quadratic Spline base} % title of Table
\centering  % used for centering table
\begin{tabular}{c|c} % centered columns (4 columns)
\hline\hline                        %inserts double horizontal lines
Type & Value \\ [0.5ex] % inserts table
%heading
\hline

Knot location, $\kappa$ & $\kappa$=(0.2, 0.4, 0.5, 0.7, 0.9) \\ % inserting body of the table
Coefficients, ${\textbf B}$ & $\beta_0 = -0.5$, $\beta_1 = 0.5$, $\beta_2 = -0.5$, and $\textbf{b} = (1,	-3,	5,	-7,	15)$ \\
Error, $\epsilon_i$ & $\epsilon_i = N(0,0.3^2)$ \\
Independent variable, $x$ & $x=(1,2,\cdots,1000)^{\rm{T}}$ \\
Dependent variable, $y$&  Sampled based on (\ref{eqn:1.1})\\[1ex]      % [1ex]
% adds vertical space
\hline %inserts single line
\end{tabular}
\label{table:nonlin2} % is used to refer this table in the text
\end{table}

The following 2 tables and 2 figures are the result of using HARARMS to find fit the free knot spline model to dataset 1 and dataset 2 correspondingly. Since the number of the knots in the true model is unknown, we fitted the models with $1, 2, \ldots, 10$ knots correspondingly. As the number of the knots increases, the maximum of the likelihood increases, even when the number is greater than the true number. However, by comparing the increasing among the different numbers of knots in the model, we can easily identify the true model. When the true number is reached, adding more knots will only increase the maximum of the log-likelihood function marginally. Of course, based on this finding, using AIC or BIC as creteria can decisively and correctly get the model selection done. The estimation of the location of the knots when the number of the knots in the model is selected correctly, is quite accurate, as we can see in the following tables.

\begin{minipage}{1.1\linewidth}
\captionof{table}{Linear regression splines} % title of Table
\label{table:nonlin3}
%\centering  % used for centering table
{\begin{tabular}{c c c c c c c c c c c c c } % centered columns (4 columns)
\hline\hline                        %inserts double horizontal lines
Case     & $\kappa_1$ & $\kappa_2$ & $\kappa_3$ & $\kappa_4$ & $\kappa_5$ & $\kappa_6$ & $\kappa_7$ & $\kappa_8$  & $\kappa_9$       &   $\kappa_{10}$ &  Log-likelihood \\ [0.5ex] % inserts table
%heading
\hline
Actual   & 200      &	300      &	400      &	500	    &  700	   & 900	  	&				    &         &         &       &        \\
\hline
1-knot   &	727 &&&&&&	&&&							&			-5631.07 \\
	5\%	        &  713		&&&&&&	&&&			&				-5633.01\\
	95\%	        & 740		&&&&&&	&&&			&						-5630.87\\
\hline
2-knot &	518	&671			&&&&&	&&&			&				-5339.78 \\
	5\%	&508	&661				&&&&&	&&&			&		-5342.61\\
	95\%	&530	&679				&&&&&	&&&			&			-5339.11\\
\hline
3-knot  &	519	&698&	899	&&&&	&&&					&			-5040.63\\
	5\%	&511	&694	&894				&&&&	&&&		&			-5043.65\\
	95\%	&526	&702	&904		&&&&	&&&				&			-5039.79\\
\hline
4-knot &	185	&540	&698	&899			&&&&	&&		&		-4925.23\\
	5\%	&172	&532&	694	&894				&&&&	&&	&		-4928.52\\
	95\%	&198	&548&	701&903			&&&&	&&			&	-4923.74\\
\hline   									
5-knot &	245	&404	&503	&697	&899		&&&	&&		&			-4802.55\\
	5\%	&232	&398	&498	&694	&896				&&&	&&	&	-4805.52\\
	95\%	&258	&409	&507	&700	&903			&&&	&&	&	-4800.69\\
\hline   												
6-knot 	&206	&286	&400	&503	&698	&899	&&&	&		&		-4796.77\\
	5\%	&168	&255	&394	&496	&695	&896		&&&	&	&		-4800.75\\
	95\%	&223	&309	&407	&507	&700	&903	&&&	&		&		-4794.36\\
\hline   												

7-knot 	 &	211	&287	&400&	503	&697	&899	&969		&&&		&-4796.57\\
	5\%	&188	&264	&394	&498	&694	&884	&899		&&&		&	-4799.86\\
	95\%	&234	&314	&406	&507	&700	&903	&987	&&&	&		-4794.46\\
\hline   												
													
8-knot 	& 	207	&227	&280	&359	&397	&503&	697	&899	&&		&-4796.20\\
	5\%	  &155&	185	&230&	313&	383&	498&	694	&895		&&		&-4799.98\\
	95\%	&224	&241	&309	&398&	410	&508	&700	&903	&&		&-4794.09\\
\hline   												
													
9-knot 	& 	30&	213	&272&	312&	396	&462	&504&	698&	899	&&-4795.85\\
	5\%	&13&	118	&210	&280	&376	&411&	498	&694	&896	&	&-4799.03\\
	95\%&	86	&231	&311&	368&	405&	507	&546&	701	&903		&	&-4793.67\\
\hline   												
10-knot &	27	&179	&218	&277	&313	&397	&474	&505	&697	&899&		-4795.03\\
	5\%	&12&	147&	182&	232&	271&	370&	393&	499&	695&	895		&-4797.91\\
	95\%	&61&	222&	279&	305&	365&	407&	508&	550&	700&	903		&-4793.02\\
\hline   												

\end{tabular}}
{The generated data is following a linear spline function with 6 knots
at 200, 300, 400, 500, 700, and 900. The $n$-th big row shows the estimation 
for the location of the knots, corresponding to the fitted model has $n$ knots. 
Also, the $90\%$ 
credible intervals to these estimation are reported. In the
last column, the value of the log-likelihood function corresponding to the 
estimated values and the boundaries of the intervals are listed. Not
surprisingly,  the log-likelihood function value of the estimations increases 
as the number of the knots increases. However, when the number of the knots in 
the model reaches the truth, the increasing in log-likelihood could be
neglected,  compared to the increasing before the number of knots in the  model
reaches the true number.}
 % is used to refer this table in the text
\end{minipage}

\begin{minipage}{1.2\linewidth}
\captionof{table}{Quadratic regression splines} % title of Table
%\centering  % used for centering table
{\begin{tabular}{c c c c c c c c c c c c c } % centered columns (4 columns)
\hline\hline                        %inserts double horizontal lines
Case     & $\kappa_1$ & $\kappa_2$ & $\kappa_3$ & $\kappa_4$ & $\kappa_5$ & $\kappa_6$ & $\kappa_7$ & $\kappa_8$  & $\kappa_9$       &   $\kappa_{10}$ &  Log-likelihood \\ [0.5ex] % inserts table
%heading
\hline
Actual         &	0.2	            &0.4              &	0.5        &	0.7    &	0.9		  	&			&&&&\\
\hline
1-knot         &	0.77	          &				         &&&&         &&&                   &                                                      			&		-1853.74   \\
5\%           &	0.77						&			             &&&&         &&&                   &                                               			&-1854.35\\
95\%	        &0.77							&		               &&&&         &&&                   &                                                      			&-1853.35\\
\hline
2-knot&	0.59&	0.64			&&&&         &&&                   &                                                      			&						-1616.39\\
5\%&	0.57&	0.62				&&&&         &&&                   &                                                      			&					-1618.44\\
95\%	&0.61	&0.65				&&&&         &&&                   &                                                      			&					-1615.81\\
\hline

3-knot	&0.58	&0.69	&0.90			&&&         &&&                   &                                                      			&					-1276.31\\
5\%	&0.57	&0.68	&0.90					&&&         &&&                   &                                                      			&			-1279.13\\
95\%	&0.59&	0.70&	0.91			&&&         &&&                   &                                                      			&					-1275.32\\
\hline

4-knot	&0.45	&0.48	&0.70&	0.90			&&         &&&                   &                                                      			&				-1205.32\\
5\%	&0.43	&0.47	&0.69	&0.90						&&         &&&                   &                                                      			&	-1207.94\\
95\%	&0.47	&0.50	&0.71	&0.91					&&         &&&                   &                                                      			&		-1203.96\\
\hline

5-knot	&0.19	&0.40	&0.49&	0.70	&0.90			&         &&&                   &                                                      			&				-1184.38\\
5\%	&0.15	&0.36&	0.47	&0.69	&0.90					&         &&&                   &                                                      			&		-1187.97\\
95\%	&0.24	&0.44&	0.51	&0.71	&0.91				&         &&&                   &                                                      			&			-1182.81\\
\hline
6-knot&	0.19&	0.41&	0.49&	0.69	&0.72	&0.90			&&&                   &                                                      			&		-1184.09\\
5\%	&0.14&	0.38&	0.47&	0.61	&0.69	&0.90				&&&                   &                                                      			&	-1187.57\\
95\%	&0.23&	0.44&	0.51&	0.71	&0.79	&0.91			&&&                   &                                                      			&		-1182.59\\
\hline

7-knot&	0.20&	0.39	&0.47	&0.51	&0.70	&0.86	&0.90	&&                   &                                                      			&			-1183.19\\
5\%	&0.15&	0.27&	0.41&	0.47&	0.67	&0.69	&0.89			&&                   &                                                      			&	-1186.42\\
95\%	&0.26	&0.44&	0.51&	0.70	&0.75&	0.91&	0.96		&&                   &                                                      			&		-1181.28\\
\hline

8-knot&	0.10	&0.20&	0.41&	0.48&	0.55&	0.68&	0.74	&0.90		&                   &                                                      			&	-1182.46\\
5\%	&0.01	&0.02	&0.15	&0.39&	0.48&	0.51&	0.68	&0.89			&                   &                                                      			&-1185.40\\
95\%	&0.25	&0.40	&0.47	&0.51	&0.71	&0.74	&0.91&	0.96		&                   &                                                      			&	-1180.74\\
\hline

9-knot&	0.19&	0.31	&0.43	&0.48	&0.60	&0.68&	0.72&	0.87&	0.91	&                                                      			&		-1182.08\\
5\%	&0.04	&0.14	&0.29	&0.39	&0.47	&0.52	&0.64	&0.69	&0.89		&                                                      			&	-1185.53\\
95\%	&0.26&	0.43	&0.51	&0.60	&0.69	&0.72	&0.90&	0.92&	0.98		&                                                      			&	-1180.18\\
\hline

10-knot&	0.06&	0.18	&0.28	&0.40&	0.47	&0.52	&0.64	&0.69&	0.77&	0.90	&-1181.09\\
5\%	&0.02	&0.06&	0.15&	0.29	&0.36&	0.44&	0.48&	0.63	&0.68	&0.89	&-1184.02\\
95\%	&0.24	&0.33&	0.42&	0.50&	0.54&	0.66	&0.71	&0.78&	0.91&	0.94	&-1179.21\\
\hline
\end{tabular}}
{The generated data is following a quadratic spline function with 6 knots at 0.2, 0.4, 0.5, 0.7  and 0.9. The $n$-th big row shows the estimation for the location of the knots, corresponding to the fitted model has $n$ knots. Also, the 90\% credible intervals to these estimation are reported. In the last column, the value of the log-likelihood function corresponding to the estimated values and the boundaries of the intervals are listed. Not surprisingly, the log-likelihood function value of the estimations increases as the number of the knots increases. However, when the number of the knots in the model reaches the truth, the increasing in log-likelihood could be neglected, compared to the increasing before the number of knots in the model reaches the true number.}
\label{table:nonlin4} % is used to refer this table in the text
\end{minipage}

%\section{ARMS algorithm and its extension}\label{sec:sec2}

%\section{Simulation Study}\label{sec:sec4}

\pagebreak\clearpage\newpage \thispagestyle{empty}

\end{document}